\documentclass[prl,twocolumn,showpacs,preprintnumbers,amsmath,amssymb]{revtex4}
\usepackage{graphicx}
\usepackage{dcolumn}
\usepackage{bm}
\begin{document}

\title{Fragmentation and the Bose-glass phase transition of the disordered 1-D Bose gas}

\author{Luca Fontanesi}
\email{luca.fontanesi@epfl.ch}
\affiliation{Institute of Theoretical Physics, Ecole Polytechnique F\'ed\'erale de Lausanne EPFL, CH-1015 Lausanne, Switzerland}
\author{Michiel Wouters}
\affiliation{Institute of Theoretical Physics, Ecole Polytechnique F\'ed\'erale de Lausanne EPFL, CH-1015 Lausanne, Switzerland}
\author{Vincenzo Savona}
\affiliation{Institute of Theoretical Physics, Ecole Polytechnique F\'ed\'erale de Lausanne EPFL, CH-1015 Lausanne, Switzerland}

\date{\today}

\begin{abstract}
We investigate the superfluid-insulator quantum phase transition in a disordered 1D Bose gas in the mean field limit, by studying the probability distribution of the density. The superfluid phase is characterized by a vanishing probability to have zero density, whereas a nonzero probability marks the insulator phase. This relation is derived analytically, and confirmed by a numerical study. This fragmentation criterion is particularly suited for detecting the phase transition in experiments. When a harmonic trap is included, the transition to the insulating phase can be extracted from the statistics of the local density distribution.
\end{abstract}

\maketitle
A disordered low-dimensional non-interacting system is expected to always be in an insulating state~\cite{abrahams}. Many-body interactions, in the case of Bose particles, may induce a phase transition to a superfluid state. This transition is conventionally defined through the superfluid fraction or by the long-range behaviour of the one-body density matrix~\cite{fisher_superfluidinsulator}. In 1D at zero temperature, this latter is expected to decay algebraically in the superfluid (SF) phase, and exponentially in the Bose-glass (BG) phase. The versatility and tunability of ultracold atomic systems have motivated, in recent years, the study of this fundamental phenomenon in a low-dimensional disordered Bose gas~\cite{aspect_anderson,inguscio_anderson,deissler_delocalization,demarco2010,deissler_long,modugno_report,sanchez_lewenstein}, for which however neither the superfluid fraction nor the one-body density matrix are easily accessible in experiments. Several different ways of characterizing this phase transition have thus recently been discussed~\cite{zoller,fontanesi,fontanesi2,delande,carrasquilla}. In spite of the many experimental attempts~\cite{inguscio_boseglass,hulet,demarco,deissler_delocalization} however, a clear evidence of the superfluid-insulator transition in 1D Bose gases has not yet been obtained.

The notion of \emph{fragmentation} has been widely used in relation to Bose-Einstein condensation~\cite{nozieres,lugan_phase}. In the case of a disordered Bose gas, fragmentation has been frequently evoked as a criterion for the transition from superfluid to Bose glass phase \cite{lugan_phase,deissler_delocalization}. To our knowledge, however, a rigorous definition of fragmentation of the density profile, and a proof of its relation to the quantum phase of the gas, are still lacking.

A quantity that naturally characterizes fragmentation is the probability distribution of the density (PDD), i.e. $P(\rho_0)$, in the limit $\rho_0\to 0$. We define a state to be fragmented if $P(\rho_0\to0)$ is nonzero. In this Letter we show that in the weakly interacting regime the density distribution fragments at the transition from superfluid to Bose glass. This result bears a clear experimental advantage, as the density profile of a Bose gas is a quantity that can be easily investigated, within the spatial resolution of the experimental apparatus. To make a link to current experiments, a realistic situation of a gas confined in a harmonic trap is examined \cite{hulet_transport}. In this case, the transition between a quasi-condensate and an insulator -- and a possible spatial separation between the two phases -- can be unveiled through a local investigation of the statistical distribution of the density profile.

To establish a link between the PDD and the phase transition, we study the superfluid fraction of the Bose gas. This quantity can be characterized by evaluating the response of the system to a velocity field, that is equivalent to imposing twisted boundary conditions~\cite{lieb_2002}. In fact, the superfluid fraction is proportional to the difference of the energies in the moving frame, $E_\Theta$, and in the rest frame, $E_0$, as~\cite{fisher_helicity}
\begin{equation}
 f_S=\frac{2mL^2}{\hbar^2N}\lim_{\Theta\to 0}\frac{E_\Theta-E_0}{\Theta^2}.
\end{equation}
Here $\Theta$ is the total phase twist, $m$ is the mass of the bosons, $L$ is the length of the system and $N$ is the number of particles. A mean field model of the weakly interacting 1D Bose gas requires a description in terms of a a density, $\hat\rho=\rho_0+\delta \hat\rho$, and a phase operator, $\hat\theta$, as $\hat\Psi(r)\simeq e^{i\hat\theta(x)}\sqrt{\hat\rho(x)}$. The validity of the Bogoliubov prescription, $\langle\delta\hat\rho\rangle/\rho_0\ll 1$, ensures that $\rho_0$ contains the relevant information about the density distribution. The ground state density is the solution of a Gross-Pitaevskii equation (GPE)~\cite{mora}
\begin{equation}
\left[-\hbar^2\partial_x^2/(2m)+V(x)+g\rho_0(x)\right]\sqrt{\rho_0(x)}=\mu\sqrt{\rho_0(x)},
\label{GPE}
\end{equation}
where $\mu$ is the chemical potential, $V(x)$ is the external potential and $g$ is the interaction constant. From the Gross-Pitaevskii energy functional, computed at leading order in the phase twist, the energy difference is given by the kinetic term
\begin{equation}
E_\Theta-E_0=\int \frac{(\nabla \theta(x))^2}{2m}\rho_0(x) \mathrm{d}x,
\end{equation}
where $\theta=\langle\hat\theta\rangle$. Minimizing this energy with the constraint $\int\nabla\theta = \Theta$ shows that the total superfluid fraction is related to the harmonic average of the density \cite{fontanesi2,altman_10} as
\begin{equation}\label{eq:fragmentation}
\frac{1}{\rho_S}=\int\frac{1}{\rho_0(x)}\mathrm{d}x=\int\frac{1}{\rho_0}P(\rho_0)\mathrm{d}\rho_0.
\end{equation}
The convergence of the integral in Eq. (\ref{eq:fragmentation}) is determined by the behavior of $P(\rho_0)$ in the limit $\rho_0\to0$. If we express $P(\rho_0\to0) = \rho_0^\beta$, then the condition to be in the superfluid phase is $\beta>0$. A nonzero value of $P(0)$ on the contrary, implies the insulator phase.

We compute the PDD numerically, by solving Eq. (\ref{GPE}) on finite size systems with periodic boundary conditions. Configuration average has been adopted to increase the precision of the statistical sampling. This analysis is done for Gauss-distributed and Gauss-correlated disorder, described by
\begin{equation}\label{eq:correlation}
\langle V(x)V(x')\rangle=\Delta_g^2 e^{-\frac{(x-x')^2}{2\eta_g^2}},
\end{equation}
where $\Delta_g$ is the disorder amplitude and $\eta_g$ is the spatial correlation length that introduces an additional energy scale, $E_c=\frac{\hbar^2}{2m\eta^2}$. The third energy entering the problem is the interaction energy, $U=g N_0/L$, with $N_0$ the number of bosons in the ground state.

\begin{figure}
\includegraphics[width=.5\textwidth]{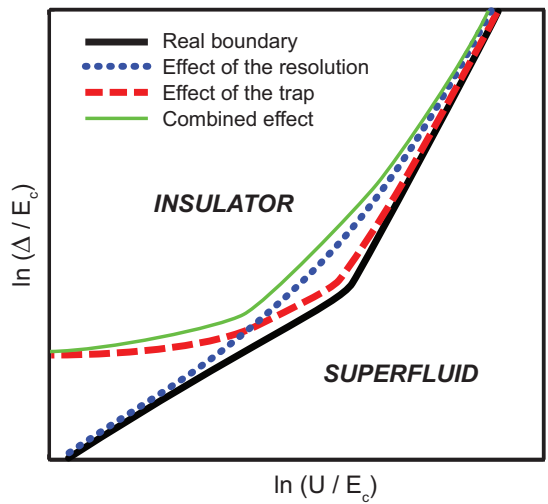}
\caption{(colors online) Phase diagram of the 1D Bose gas in presence of disorder. The effect of realistic limitations in experiments are qualitatively shown. The black thick line represents the actual phase boundary, the blue dotted line is the phase boundary extracted with a finite resolution of the experimental setup, for a fixed ratio $\eta/R$ (see text). The red dashed line is the fragmentation boundary considering a harmonic potential of fixed trapping frequency. The trap tends to displace the boudary toward the insulating phase. For $\hbar \omega_t \gtrsim U$ the local density approximation breaks down and the profile is never fragmented. The combined effect is shown by the green thin line.}
\label{figure_qualitative_pd}
\end{figure}

Previously we have characterized the phase boundary in independent ways, through the study of the superfluid fraction and the one-body density matrix~\cite{fontanesi,fontanesi2}. Two limiting cases have been identified~\cite{fontanesi2}: a Thomas-Fermi regime (TF), where $E_c\ll U$, and a white-noise limit (WN), marked by $E_c\gg U$.  In these regimes, the phase boundary obeys power-law relations, $\Delta/E_c=C(U/E_c)^\gamma$, with $\gamma$ respectively equal to $1$ and $3/4$ (cfr Fig. \ref{figure_qualitative_pd}).

\begin{figure}
\includegraphics[width=.5\textwidth]{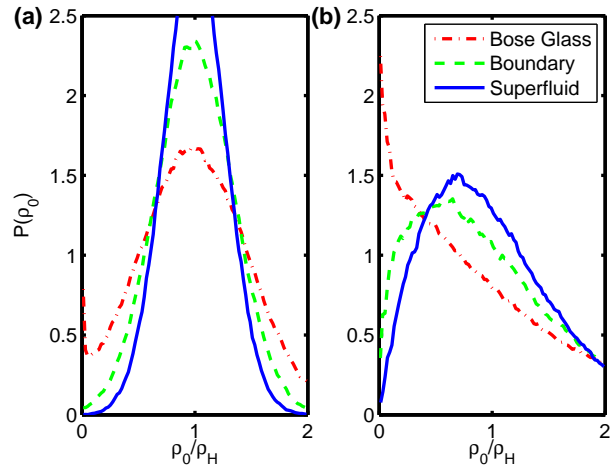}
\caption{(color online) Probability distribution of the density in two different regimes. $(a)$ Towards the TF regime, $\Delta_g = 12.8 \; E_c$ for $U= 25.6\; E_c $ (red dot-dashed)$,\;35.84 \; E_c $ (green dashed)$,\; 46\; E_c$ (blue solid). $(b)$ Towards the WN limit, $\Delta_g = 0.016 \; E_c$ for $U= 0.0032\; E_c$ (red dot-dashed)$,\;0.0048 \; E_c $ (green dashed)$,\; 0.0064\; E_c$ (blue solid). $\rho_H$ is a constant value, solution of the homogeneous case.}
\label{figure_fragmentation}
\end{figure}

Fig. \ref{figure_fragmentation}$(a)$ shows the PDD for fixed $\Delta_g=12.8 \; E_c$ and increasing interaction $U= 25.6\; E_c,\;35.84 \; E_c,\; 46.08\; E_c$ (all in the TF regime). From our previous analysis~\cite{fontanesi,fontanesi2} these three values correspond to the BG, phase boundary and SF phases respectively. In the homogeneous case $P(\rho_0)$ is expected to have a single peak at the value $\rho_0/\rho_H=1$, where $\rho_H$ is the constant solution of the homogeneous problem. The inclusion of a small disorder~\cite{sanchez_smoothing} broadens this peak, but $P(\rho_0)$ preserves a vanishing tail for $\rho_0\to 0$ (solid blue curve in Fig. \ref{figure_fragmentation}$(a)$) in the superfluid phase. For increasing disorder the weight of the low-density part becomes more important (green dashed curve in Fig. \ref{figure_fragmentation}$(a)$) until the phase boundary is eventually crossed and the PDD develops a finite component in the limit $\rho_0\to 0$ (red dot-dashed line in Fig. \ref{figure_fragmentation}$(a)$). Fig. \ref{figure_fragmentation}$(b)$ shows a similar analysis carried out for the WN regime, for $\Delta_g=0.016 \; E_c$ and $U= 0.0032\; E_c,\;0.0048 \; E_c,\; 0.0064\; \; E_c$. As for the TF case, from our previous study these three values lie in the BG, phase boundary and SF phases respectively.

Comparing panels $(a)$ and $(b)$ in Fig. \ref{figure_fragmentation}, it is clear that the PDD has different shapes in the TF and WN regimes. However, in both cases, the fragmentation allows to differentiate between SF and BG phases.
The numerical analysis summarized in Fig. \ref{figure_fragmentation} confirms the criterion stemming from eq. (\ref{eq:fragmentation}), namely that the SF fraction is nonzero if and only if $P(\rho_0\to0)=0$. We have checked that the same conclusions hold also for a speckle potential. Again, the choice of the potential strongly affects the shape of the PDD, but the limiting behaviour $P(\rho_0\to0)$ is only determined by the phase of the gas.

$P(\rho_0)$ can be evaluated analytically in certain regimes. In the TF regime, when the kinetic term is negligible, the density follows the external potential according to the TF approximation as
\begin{eqnarray}\label{eq:TF}
 \rho_0(x) = & [\mu-V(x)]/g \qquad & \textrm{if} \qquad V(x)<\mu, \nonumber \\
 \rho_0(x) = & 0 \qquad & \textrm{if} \qquad V(x)>\mu.
\end{eqnarray}
In this regime the distribution of $\rho_0(x)$ reproduces the distribution of the potential at any finite value, with an additional finite contribution in zero, given by the sum of the regions where $V(x)>\mu$. A very similar feature can be indeed noticed in the insulating case of Fig. \ref{figure_fragmentation}$(a)$ (dot-dashed line) where the PPD has a gaussian-like shape with a peaked contribution in $0$. Following the fragmentation argument, this case is always insulating as it can be expected since the absence of a kinetic component prevents the formation of any quasi-long range order or superfluid flow. This is not in contradiction with the phase transition found for large values of $U/E_c$, because the kinetic energy corrections to (\ref{eq:TF}) are responsible for the build-up of the quasi-long-range order.

Homogeneous systems are useful theoretical tools to inspect the phase transition, but current experiments \cite{hulet,deissler_delocalization} are performed on trapped systems. The harmonic trap introduces a spatial inhomogeneity in the interaction energy $U(x)= g\rho_0(x)$. To deal with a realistic situation we consider a speckle disorder on the top of an harmonic trapping potential, as it has been investigated in recent experiments~\cite{hulet,aspect_densitymodulation}, and we perform the simulation on a system of experimentally achievable size~\cite{hulet_transport} ($200$ correlation lengths). The speckle potential results from the interference pattern produced by the scattering of coherent light through a rough plate. Its amplitude, $\Delta_s$, is defined as the standard deviation of the potential and its distribution is bound from one side (above or below depending on the detuning of the laser) and decays exponentially on the opposite side~\cite{goodman}. Its correlation length $\eta_s$ is given by the numerical aperture of the focusing lens, and in particular is linked to the cutoff in \emph{k}-space $k_c$ as $\eta_s=1/k_c$.

\begin{figure}
\includegraphics[width=.5\textwidth]{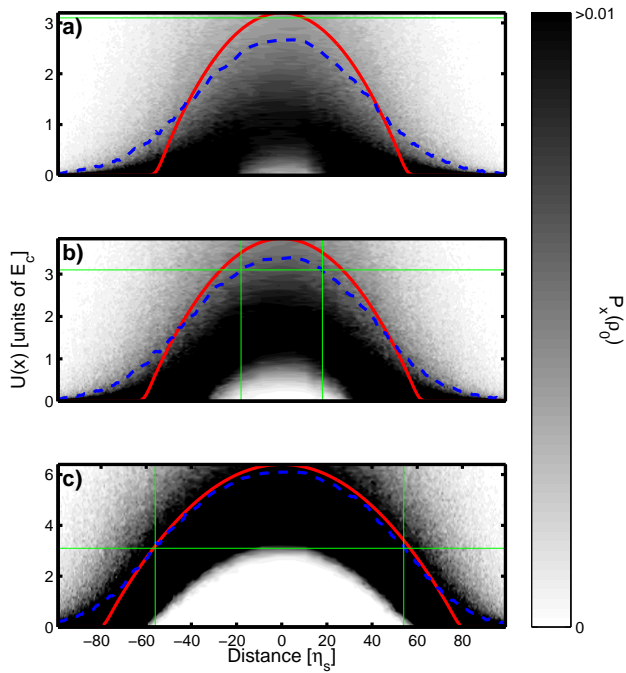}
\caption{Local probability distribution of the density in a harmonic trap ($\hbar \omega_t = \; 0.064 \; E_c$) for three different values  of the interaction energy $U_{C}(0)$: (a) $3.2 \; E_c$, (b) $3.84 \; E_c$, (c) $6.4 \; E_c$ and for fixed speckle disorder intensity ($\Delta_s= 3.2 \; E_c$). The distributions are obtained from configuration averages. The red solid lines are the interaction energy profiles in the disorderless cases. Blue dashed lines represent the spatial distribution of the average interaction energies. The green thin lines indicate the boundary in the thermodynamic limit.}
\label{figure_trap}
\end{figure}

\begin{figure}
\includegraphics[width=.5\textwidth]{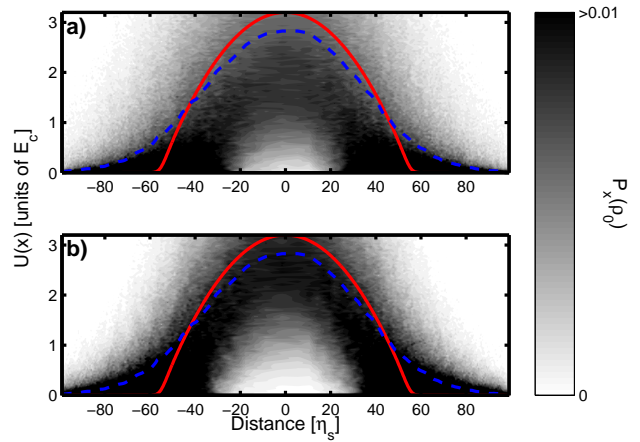}
\caption{Effect of a finite resolution on the local probability distribution of the density in a harmonic trap ($\hbar \omega_t = \; 0.064 \; E_c$) for the insulating state of Fig. \ref{figure_trap}(a): $U_{C}(0)=3.2 \; E_c$ and $\Delta_s=3.2 \; E_c$. Panel (a) shows a resolution $R=1$, whereas in panel (b) $R=2$.}
\label{figure_resolution}
\end{figure}

The contour plots in Fig. \ref{figure_trap} show the spatially resolved probability distributions. $P_x(\rho_0)$, as a function of position and interaction energy ($U(x)$) for fixed disorder amplitude, $\Delta_s = 3.2 \;E_c$. The blue dashed line is the average interaction energy $\overline{U}(x) = \int g \rho_0(x) P_x(\rho_0)\mathrm{d}\rho_0$, and the red solid line represents the interaction profile in absence of disorder $U_{C}(x)$, i.e. the solution of the GPE with $V(x)=m\omega_t^2 x^2/2$, where $\omega_t$ is the trapping frequency. As we are considering a speckle potential bound from above (attractive), the distribution $P_x(\rho_0)$ is not centered around its mean value. Note also that the disorder combined with a smooth trapping potential makes the average density profile different from the clean one, in fact it extends beyond the disorderless TF radius and it reaches a lower value in the center of the trap. This fact is relevant when comparing the critical values of interaction at the transition between the trapped and untrapped cases.

It is important to understand whether the fragmentation threshold, in the trapped case, coincides with the one extracted in the thermodynamic limit. For the chosen value of the disorder amplitude, the fragmentation analysis in the untrapped case gives a transition at an interaction strength $U_H \simeq 3.1 \; E_c$. In panel $(a)$ the gas is fragmented at all positions, consistent with the homogeneous boundary, as the maximal interaction energy $U(0)\simeq2.7\;E_c<U_H$. When the interaction energy is increased, SF and BG phases coexist (see panels $(b)$ and $(c)$). The fragmentation threshold of the interaction energy can be extracted directly from the spatially resolved PDDs, these values for the average interaction at the threshold coincide with $U_H$ (cfr. Fig. \ref{figure_trap}$(b)$ and $(c)$). This is true as long as the local density approximation is valid ($\hbar \omega_t \ll U$). For tighter traps (simulations not shown), the fragmentation line moves to lower values of the average interaction energy, that can be interpreted as a penetration of the SF into the insulating state (cfr. Fig. \ref{figure_qualitative_pd}). This is obvious when considering the limit $\hbar \omega_t \gg U,\Delta$, where the density profile is the non-fragmented harmonic oscillator ground state. The penetration effect is more pronounced when the transition occurs at the edges of the trap where the density gradient is steeper. Therefore, the best agreement between the phase transition of the trapped and homogeneous gases is obtained when the transition occurs close to the center of the trap.

In current experiments the resolution of the apparatus represents the main limitation to the correct interpretation of the experimental data. For this purpose we study the role of a finite resolution on the statistical analysis that we propose in this article. The resolution is included in the simulation via a convolution of the ground state wavefunction with a gaussian of standard deviation $R$. The finite resolution makes it harder to identify the insulator phase through a statistical study of the density because it smooths out the profile, cutting narrow minima from the statistics. This limitation acts differently in different regimes: supposing the ratio $\eta/R$ fixed, the finite resolution would not affect the statistics in the WN limit, where modulations of the density occur on the scale of the healing length, much longer than the resolution. On the contrary, if $\eta/R \lesssim 1$, this would strongly affect the statistics in the TF regime, where the lengthscale of the modulation would be the typical correlation length of the potential. Fig. \ref{figure_resolution} shows the effect of finite resolution on the probability distribution of a trapped Bose gas for $U_C(0)=3.2\;E_c$ and $\Delta_s=3.2\; E_c$. The statistical analysis of the distribution (Fig. \ref{figure_trap}(a)) identifies a completely fragmented state. The introduction of a finite resolution ($R=1$) in Fig. \ref{figure_resolution}(a) opens a window in the center of the trap as if the Bose gas was partially superfluid. This effect become even more significant for a worse resolution ($R=2$), as shown in Fig. \ref{figure_resolution}(b), where the gas appears to be predominantly superfluid. For this set of parameters, the transition would be placed at $U \sim 3,\;2.9  \; E_c$ respectively.

Both the finite resolution and the harmonic trap tend to displace the apparent fragmentation threshold towards the insulator phase with respect to the actual boundary. Therefore, in experiments, the superfluid fraction is overestimated, but a quantitative correction to the boundary can be computed knowing the value of the resolution and the intensity of the trap. A qualitative sketch of the effect of real conditions on the actual phase diagram is shown in Fig. \ref{figure_qualitative_pd}. From this sketch it appears that the trapped system does no longer reflect the properties of the thermodynamic limit deep in the WN regime. Indeed, fragmentation would occur on a scale much longer than the ground state of the trap.

In conclusion, we have demonstrated a relation between the quantum phase transition of the 1D Bose gas and the probability distribution of the density in the mean field limit. We found that the superfluid phase is marked by a vanishing probability at zero density whereas the insulating phase develops a finite component in the limit $P(\rho_0\to0)$. This gives a viable route to the determination of the phase of the gas, and therefore of the phase transition, through a statistical analysis of the density profiles.

We thank B. Deissler, R. Hulet, P. Lugan, E. Runge and L. Sanchez-Palencia for stimulating discussions. This research has been supported by the Swiss National Science Foundation through Project No. 200021-117919.

\end{document}